\documentclass[a4paper,11pt]{article}
\pdfoutput=1
\usepackage{graphicx}
\usepackage{epsf,amsmath,bbold,amsfonts,stmaryrd}

\usepackage[utf8]{inputenc}
\usepackage{mathrsfs}
\usepackage{appendix}
\usepackage{amssymb}
\usepackage{float}
\usepackage{color}
\usepackage{cite}
\usepackage{hyperref}
\hypersetup{pageanchor=false}
\usepackage{indentfirst}
\usepackage{url}
\usepackage{float}
\usepackage{caption}
\usepackage[numbers,square,comma,sort&compress,merge]{natbib}

\def\mc{\mathcal}

\def\be{\begin{equation}}
\def\ee{\end{equation}}

\def\bea{\begin{eqnarray}}
\def\eea{\end{eqnarray}}

\def\ba{\begin{array}}
\def\ea{\end{array}}

\def\bc{\begin{center}}
\def\ec{\end{center}}

\def\bl{\begin{flushleft}}
\def\el{\end{flushleft}}

\def\br{\begin{flushright}}
\def\er{\end{flushright}}

\def\bi{\begin{itemize}}
\def\ei{\end{itemize}}

\def\bt{\begin{tabular}}
\def\et{\end{tabular}}
\newcommand{\sR}{\mathsf{R}}

\begin{document}
\title{\textbf{ On AdS Black Strings at
Large $D$}}
\date{}
\author{Peng-Cheng Li$^{1}$\footnote{pchli2021@scut.edu.cn},\quad Cheng-Yong Zhang $^{2}$\footnote{ zhangcy@email.jnu.edu.cn, Corresponding author.}}

\maketitle
\begin{center}
{\it

$^1$School of Physics and Optoelectronics, South China University of Technology, Guangzhou 510641, People’s Republic of China\\\vspace{4mm}
$^2$Department of Physics and Siyuan Laboratory, Jinan University,
Guangzhou 510632, China
}
\end{center}
\begin{abstract}
In this paper we study the stability of an homogeneous black string
in the presence of a negative cosmological constant with minimally coupled scalar fields by using the large $D$ effective theory. This method allows us to explore the dynamics of the black strings in the nonlinear regime. We find that up to the next-to-leading order of the $1/D$ expansion, the unique consistent solution of the equations of motion  must be uniform. This means the recently found Gregory-Laflamme instability caused by fined-tuned non-generic perturbations would die out at late time in the evolution of the system.
\end{abstract}

\newpage
\section{Introduction}
Black holes in higher dimensions $(D\geq4)$ have much richer physics than in four dimensions \cite{Emparan:2008eg}. A typical example is black string, whose horizon topology is $S^{D-3}\times S^1$. The black strings suffer from the so-called Gregory-Laflamme (GL) instability when the wavelength of the gravitational perturbations is long, or equivalently, the black strings are thin enough \cite{Gregory:1993vy}. Numerical simulations at $D=5$ shows that the evolution of the GL instability will lead to a pinch-off of the horizon making visible the naked singularity and the violation of the weak cosmic censorship \cite{Lehner:2010pn}.

In contrast to Ricci-flat case, a uniform
black string in $D$ dimensional AdS spacetime cannot be constructed by simply adding a trivial direction to a  Schwarzschild-AdS black hole in $D-1$ dimensional spacetime. Instead, one has a warping factor in front of the metric, which makes the AdS black string to be non-uniform. In general, homogenous black strings in AdS can only be constructed numerically \cite{Copsey:2006br}. Recently, however, an homogeneous AdS black string in $D\geq4$ was constructed analytically  in the presence of minimally coupled free scalar fields \cite{Cisterna:2017qrb}. By focusing on the perturbations of the scalar fields, which are decoupled from the perturbations of the metric, Ref. \cite{Cisterna:2019scr} showed that these solutions are linearly stable. Very recently, it was found that the AdS black strings could also suffer from the GL instability in $D\geq5$, by considering  only perturbations of the metric \cite{Henriquez-Baez:2021gdn,Dhumuntarao:2021gdb}. Since this instability is achieved in the absence of the scalar perturbations, it would be interesting to ask whether such instability maintains beyond the linear regime and  if so  what could be its fate. Thus, a full nonlinear treatment for this issue is necessary.

The goal of this work is to better understand the dynamics of the newly  discovered  AdS black string solutions by employing the large $D$ expansion method \cite{Emparan:2013moa}. In the large dimension limit, a lot of non-trivial dynamics of black holes can be completely localized at the near-horizon region and then a full nonlinear effective field theory can be formulated \cite{Emparan:2015hwa}. Using this effective theory one can  construct various black hole solutions and furthermore study
their dynamics linearly or nonlinearly \cite{Suzuki:2015iha,Tanabe:2015hda,Suzuki:2015axa,Emparan2015,Andrade:2018nsz,Chen:2017hwm}. In particular, the large $D$ effective theory has been successfully applied to the dynamics of black strings \cite{Suzuki:2015axa,Emparan2015}. For example, it was shown that the end state of large $D$ unstable black strings is always  stable non-uniform black strings \cite{Emparan2015}.

In the next section we present the large $D$ study of the AdS black strings.
\section{Large $D$ study}
\subsection{Setup}
Consider Einstein theory in dimension $D=n+4$, minimally coupled to one scalar field $\psi$. The field equations are given by
\be\label{EEqs}
R_{\mu\nu}-\frac{1}{2}g_{\mu\nu}R-\frac{(D-1)(D-2)}{2\ell^2}g_{\mu\nu}=\kappa T_{\mu\nu},
\ee
with
\be
T_{\mu\nu}=\frac12 \partial_\mu\psi \partial_\nu\psi-\frac{1}{4}g_{\mu\nu}\partial_\rho\psi\partial^\rho\psi,
\ee
and
\be\label{KGeq}
\nabla_\mu\nabla^\mu\psi=0.
\ee
Hereafter, we will set $\kappa=16\pi G=1$.

Ref. \cite{Cisterna:2017qrb} found that above equations admit the following black string solution
\be
ds^2=-f(r)dt^2+\frac{dr^2}{f(r)}+dz^2+r^2d\Omega^2_{D-3},
\ee
with
\be
f(r)=1-\frac{2\mu}{r^{D-4}}+\frac{D-1}{D-2}\frac{r^2}{\ell^2},
\ee
and  the solution for the fields takes the simple form
\be\label{scalarfield}
\psi=\frac{\sqrt{2(D-1)}}{\ell} z.
\ee
Here $\mu$ is an  integration constant related to the mass of the black hole and $z$ is the coordinate along the string direction with period $L$. Note that this spacetime is asymptotically $AdS_{D-1}\times S^1$ with the curvature radius of the $AdS_{D-1}$ factor given by
\be
\frac{1}{\ell_c^2}\equiv\frac{D-1}{D-2}\frac{1}{\ell^2}.
\ee
The  dressed curvature radius $\ell_c$  equates to the bared one $\ell$  at large $D$.

\subsection{Large $D$ solutions}
Inspired by above uniform black string solution, in terms of the ingoing Eddington-Finkelstein coordinates, we make the metric ansatz for the dynamical AdS black string solution as
\be
ds^2=-Adv^2+2(u_vdv+u_zdz)dr-2C_z dzdv+G_{zz}dz^2+r^2d\Omega_{n+1}^2,
\ee
where all the functions appeared in the metric depend on $(v,r,z)$.

The ansatz for the scalar field is just
\be
\psi=\psi(v,r,z).
\ee
In order to perform the $1/n$ expansion properly we need to specify the large $n$ behaviors of
the metric functions and the scalar field. First of all, we note that the left hand side of the Einstein equations (\ref{EEqs}) is of $\mc O(n^2)$, so the stress-tensor of the scalar field is of the same order only if we rescale $z\to z/\sqrt{n}$. Besides, the zero-mode wavenumber of the  GL instability  of the Ricci-flat black strings  scales like $k_{GL}\simeq\sqrt{n}/r_0$ with $r_0$ being the thickness of the black strings \cite{Emparan:2013moa}, which indicates the same rescaling should be done to capture the unstable fluctuations along the string direction. We expect that the critical wavenumber of the AdS black strings is of the same order. In addition, we consider a small velocity $\mc O(1/\sqrt{n})$ along the string direction. Thus, the large $n$ scalings of the metric functions inferred from the exact solution are given by
\be
A=\mc O(1),\quad u_v=\mc O(1),\quad u_z=\mc O(n^{-1}),\quad C_z=\mc O(n^{-1}),\quad G_{zz}=\frac{1}{n}\left(1+\mc O(n^{-1})\right),
\ee
and the scalar field
\be
\psi=\mc O(1).
\ee
By a gauge choice we can set $u_z=0$. Thus, we have
\be
A=A^{(0)}+\frac{A^{(1)}}{n}+\cdots,\quad u_v=u_v^{(0)}+\frac{u_v^{(1)}}{n}+\cdots, \quad C_z=\frac{C_z^{(0)}}{n}+\frac{C_z^{(1)}}{n^2}+\cdots,
\ee
and
\be
G_{zz}=\frac{1}{n}\left(1+\frac{G_{zz}^{(0)}}{n}+\cdots\right),\quad \psi=\psi^{(0)}+\frac{\psi^{(1)}}{n}+\cdots.
\ee
At large $n$ the radial gradient becomes dominant, that is $\partial_r=\mc O(n)$, $\partial_t=\mc O(1)$, $\partial_z=\mc O(1)$, so
in the near-horizon region of the black hole it is better to use a new radial coordinate $\sR$ defined by
\be\label{newradial}
\sR=\left(\frac{r}{r_0}\right)^n,
\ee
such that $\partial_\sR=\mc O(1)$, where $r_0$ is a horizon length scale which can be set to unity $r_0=1$.
To solve the field equations we need to specify boundary conditions for the metric functions at large $\sR$, they are given by
\be\label{bdy}
A=1+\frac{1}{\ell^2}+\mc O(\sR^{-1}),\quad u_v=1+\mc O(\sR^{-1}), \quad C_z=\mc O(\sR^{-1}),\quad G_{zz}=\frac{1}{n}\left(1+\mc O(\sR^{-1})\right).
\ee
That is to say  the solution should asymptotically approach to $AdS_{n+3}\times S^1$. For the scalar field, we require it approaches to the exact solution (\ref{scalarfield}) as $\sR\to\infty$, since the exact black string solution is still valid in that regime. Thus, we may impose the boundary condition for the scalar field as
\be\label{bdyscalar}
\psi=\frac{\sqrt{2}}{\ell}z+\mc O(\sR^{-1}),
\ee
at large $\sR$.
   On the other hand, the solutions have to be regular at the horizon.

At the leading order of the $1/n$ expansion, the field equations only contain $\sR$-derivatives so
they can be solved by performing $\sR$-integrations. Then after imposing the boundary conditions the
leading order solutions are obtained as
\be
A^{(0)}=1+\frac{1}{\ell^2}-\frac{m(v,z)}{\sR},\quad u_v^{(0)}=1,\quad C_z^{(0)}=\frac{p(v,z)}{m(v,z)\sR},
\ee
\be\label{LOSol2}
G_{zz}^{(0)}=\frac{p(v,z)^2}{m(v,z)\sR},\quad \psi^{(0)}=\frac{\sqrt{2} }{\ell}z.
\ee
In the above expressions, $m(t,z)$ and $p(t,z)$ are the integration functions
of $\sR$-integrations of the field equations, which can be viewed as the mass and momentum
density of the solution. Obviously, the uniform black string corresponds to $m(v,z)={\rm constant}$ and $p(v,z)=0$.
Surprisingly, we find that the leading order of the scalar field is just the same as that of the exact solution (\ref{scalarfield}) at large $n$. This can be explained as follows. The $rr$ component of the Einstein equations at the leading order of the $1/n$ expansion tells us that the scalar field is independent of the radial coordinate, i.e., $\psi^{(0)}=\psi^{(0)}(v,z)$. Then we can find that the scalar field appears in the expression of $G_{zz}^{(0)}$,
\be
G_{zz}^{(0)}=\frac{(2-\ell^2  (\partial_z\psi^{(0)})^2)\log \sR}{\ell^2+1}+\frac{p^2}{\sR m},
\ee
which is in conflict with the boundary conditions (\ref{bdy}), unless the scalar field is simply given by (\ref{LOSol2}).
In fact, once we know the leading order scalar field is independent of $\sR$, the boundary condition (\ref{bdyscalar}) tell us that the solution has to be (\ref{LOSol2}).

At the next-to-leading order of the $1/n$ expansion, we find something strange. The Klein-Gordon equation (\ref{KGeq}) reduces to
\be
 \left(-\ell^2 m+\ell^2 \sR+\sR\right)\frac{d^2\psi^{(1)}}{d\sR^2}+\left(\ell^2+1\right) \frac{d\psi^{(1)}}{d\sR}-\frac{\sqrt{2} \ell p}{\sR^2}=0,
\ee
where  $\psi^{(1)}$  denotes the subleading order of the scalar field in the expansion of $1/n$.
 Take into account the boundary conditions at infinity (\ref{bdyscalar}), we obtain the solution
 \be
\psi^{(1)}=\frac{3z}{\sqrt{2}\ell}-\frac{\sqrt{2} p}{\ell m}\log\left(1-\frac{\ell^2}{1+\ell^2}\frac{m}{\sR}\right).
 \ee
We can observe that this solution becomes divergent at the horizon $\sR=\frac{\ell^2}{\ell^2+1}m$, unless $p=0$. Moreover, provided $p=0$, from the $vv$ and $vr$  components of the Einstein equations we get an equation that $m$ have to satisfy
\be
\partial_v m-\partial_z^2m=0,
\ee
and from the $vz$ component we have
\be
\partial_z m=0.
\ee
Then the unique solution is  $m={\rm constant}$, which means the AdS black string solution has to be uniform. In the Ricci-flat case \cite{Emparan2015}, the counterparts of above two equations are the effective equations encoding the non-trivial dynamics of black strings at large $D$.
\section{Final remark}
We studied  the recently discovered  AdS black string solutions by employing the large $D$ effective theory. This method allows us to explore the dynamics of the black strings in the nonlinear regime. Expanding the Einstein and Klein-Gordon equations up to the next-to-leading order in $1/n$, we find that the unique consistent solution to these equations is the uniform black string, which means the GL instability due to the metric perturbations and in the absence of the scalar perturbations cannot maintain to the nonlinear regime. Thus, we expect that even the initial uniform black string becomes unstable due to this fine-tuned non-generic perturbations, at late time of evolution the end state will be a uniform black string. This result is also in accord with the analysis in \cite{Cisterna:2019scr} that the AdS black string is linearly stable under generic perturbations.

It worth noticing that the large $D$ effective theory only applies to dynamics of black holes that is decoupled  from  the asymptotically far region,
so the validness of the discussion in the work relies on the assumption that the dynamics of the AdS black string is of this kind. However, this is not guaranteed in priori, despite the whole analysis is self-consistent. Therefore a complete numerical study for the nonlinear dynamics of the AdS black string in arbitrary $D$ can give more convincing judgement.

\section*{Acknowledgments}
This research is supported by the Natural Science Foundation of China under Grant No. 12005077  and Guangdong Basic and Applied Basic Research Foundation under Grant No. 2021A15 15012374.

\end{document}